\documentclass[ superscriptaddress, aps, pre, twocolumn, floatfix ]{revtex4-2}

\usepackage[ usenames, dvipsnames ]{xcolor}
\usepackage{amsmath, amsfonts, amssymb, amsthm, amstext, bm, bbm}
\usepackage[colorlinks=true, linkcolor=blue,citecolor=red, urlcolor=blue, hypertexnames=true]{hyperref}
\usepackage{graphicx, subcaption}

\usepackage{ragged2e}
\DeclareCaptionJustification{justified}{\justifying}
\newcommand{\brac}[1]{\left( #1 \right) }

\usepackage[normalem]{ulem}

\begin{document}

\title{Out-of-time-ordered correlator in the one-dimensional Kuramoto-Sivashinsky and Kardar-Parisi-Zhang equations}

\author{Dipankar Roy}
\email[ ]{dipankar.roy@icts.res.in}
\affiliation{International Centre for Theoretical Sciences, Tata Institute of Fundamental Research,
Bangalore 560089, India}

\author{David A. Huse}
\email[ ]{huse@princeton.edu}
\affiliation{Physics Department, Princeton University, Princeton, NJ, 08544, USA}

\author{Manas Kulkarni}
\email[ ]{manas.kulkarni@icts.res.in}
\affiliation{International Centre for Theoretical Sciences, Tata Institute of Fundamental Research,
Bangalore 560089, India}

\date{\today}

\begin{abstract}
The out-of-time-ordered correlator (OTOC) has emerged as an interesting object in both classical and quantum systems for probing the spatial spread and temporal growth of initially local perturbations in spatially extended chaotic systems. Here, we study the (classical) OTOC and its ``light-cone'' in the nonlinear Kuramoto-Sivashinsky (KS) equation, using extensive numerical simulations.  We also show that the {\it linearized} KS equation exhibits a qualitatively similar OTOC and light-cone, which can be understood via a saddle-point analysis of the linearly unstable modes. Given the deep connection between the KS (deterministic) and the Kardar-Parisi-Zhang (KPZ which is stochastic) equations, we also explore the OTOC in the KPZ equation. While our numerical results in the KS case are expected to hold in the continuum limit, for the KPZ case it is valid in a discretized version of the KPZ equation. More broadly, our work unravels the intrinsic interplay between noise/instability, nonlinearity and dissipation in partial differential equations (deterministic or stochastic) through the lens of OTOC.
\end{abstract}

\maketitle

\textit{Introduction. --} The spatiotemporal spread of perturbations is a topic of great interest in spatially extended, chaotic systems. The \emph{out-of-time-ordered correlator} (OTOC) has been recently proposed as a diagnostic tool to understand the growth (or decay) of perturbations in such systems. The OTOC captures both the temporal growth and the spatial spread of an initially localized  
perturbation. This quantity has been used in classical models, in particular models or systems which involve a large number of degrees of freedom.  In classical systems, the initial local perturbation can be infinitesimal.  For example, the OTOC has been used to study spreading of perturbations in a classical spin chain \cite{2018-das--bhattacharjee}, chaos in thermalized fluids \cite{2021-murugan--ray}, many-body chaos in classical interacting spins on a kagome lattice \cite{2018-bilitweski--moessner, 2021-bilitewski--moessner}, classical disordered anharmonic chain \cite{2020-kumar--dhar}, chaos and anomalous diffusion across a thermal phase transitions in 2D XXZ model with anisotropy \cite{2021-ruidas-banerjee}, dynamical regimes of ﬁnite temperature discrete nonlinear Schr\"{o}dinger chain \cite{2021-chatterjee--kundu}, driven dissipative Duffing chain \cite{2020-chatterjee--kulkarni}, low dimensional classical chaotic systems \cite{2018-jalbert--wisniacki}, velocity dependent Lyapunov exponents in classical chaos \cite{2018-khemani--nahum}, power-law models at low temperatures \cite{2021-s--kulkarni}, integrable spin chains including effects of breaking integrability \cite{2022-roy--kulkarni}, coupled map lattices \cite{muruganandam2022manifestation} and spin chains with kinetic constraints \cite{2022-deger--lazarides, 2022-deger--roy}.

Despite these extensive studies, the OTOC in continuum nonintegrable (in the Liouville sense) systems or nonintegrable partial differential equations has not received much attention.  In addition, much remains to be explored regarding the intrinsic interplay between instability, dissipation and nonlinearity. In this context, a natural candidate is the well-known Kuramoto-Sivashinsky (KS) equation \cite{1976-kuramoto-tsuzuki, 1977-sivashinsky}. This is a deterministic equation where there is a rich interplay between instability and chaos that leads to an emergent noise, provided there are sufficient number of unstable modes, which happens in the limit of large system size \cite{1989-zaleski, 1993-hayot--josserand}. On the other hand, certain aspects, such as scaling, spatiotemporal correlations, and distributions of height fluctuations of the well-known Kardar-Parisi-Zhang (KPZ) equation \cite{1986-kardar--zhang, 1995-healy-zhang, 2018-takeuchi} are deeply connected to the KS equation \cite{1992-sneppen--bohr, 1993-hayot--josserand, 2020-roy-pandit}. This naturally raises the question of the possible connection between these two models as far as OTOC is concerned.


\textit{Models and properties. --} We will start by discussing some relevant details of  the KS and KPZ equations. The KS equation \cite{1975-kuramoto-tsuzuki, 1976-kuramoto-tsuzuki} reads
\begin{equation}
    \partial_{t} h = -  \partial^{2}_{x} h -  \partial^{4}_{x} h  -   \frac{ 1 }{2} (\partial_{x} h )^{2} 
    \label{eq:1dks}
\end{equation}
where $h(x,t)$ is a height profile defined on $x \in [0, L]$ with periodic boundary conditions. The KS equation embodies an intriguing interplay of instability, dissipation, and nonlinearity represented by the first, second and the third term respectively in the right-hand-side of Eq.~\eqref{eq:1dks} \cite{1992-sneppen--bohr, 1993-hayot--josserand}. The KS equation appears in various physical contexts, such as propagation of waves in dissipative media \cite{1975-kuramoto-tsuzuki, 1976-kuramoto-tsuzuki}, flame front propagation \cite{1977-sivashinsky, 1980-sivashinsky}, diffusion-induced chemical turbulence \cite{1980-kuramoto}, irregular flow of liquid film down a vertical plane \cite{1980-sivashinsky-michelson, 1982-shlang-sivashinsky, 1983-pumir--pomeau}, model system with intrinsic stochasticity \cite{1984-pomeau--pelce}, dynamical systems \cite{1985-nicolaenko--temam, 1986-hyman-nicolaenko}, and phase turbulence \cite{1978-kuramoto, 1996-grinstein--pandit}, to name a few. Besides its importance in modelling diverse physical phenomena, the 1D KS equation has an interesting connection with the 1D KPZ equation \cite{1986-kardar--zhang}, a typical model under the KPZ universality class \cite{1995-healy-zhang, 2015-healy-takeuchi, 2018-takeuchi}. Numerical and theoretical investigations \cite{1992-sneppen--bohr, 1992-lvov-procaccia, 1993-lvov--procaccia, 1993-hayot--josserand, 1995-chow-hwa, 2020-roy-pandit} suggest that the long-time and large-length properties of the KS equation correspond to those of the KPZ equation. This deep connection is rooted in the unstable long-wavelength modes and the spatiotemporal chaos in the KS equation, which are responsible for generating an effective ``noise''.

The 1D KPZ equation is given by
\begin{equation}
	\partial_{t} h  =  \partial^{2}_{x} h + g (\partial_{x} h )^{2} + \eta ,
	\label{eq:1dkpz}
\end{equation}
where $h(x,t)$ is the fluctuating and growing height field, $g$ is the strength of nonlinearity and $\eta$ is the Gaussian white noise with strength 1: 
\begin{equation}
	\langle \eta(x,t) \eta(x',t') \rangle = 2 \delta(x-x') \delta(t-t'). 
\end{equation}
Note that the parameter $g$ is taken to be $g=8$ for numerical convenience. However, by suitably scaling the space, time, and the height field, the parameter $g$ can also be set to $1$ in Eq.~\eqref{eq:1dkpz} \cite{2012-corwin}.


In this paper we study how localized perturbations behave in these two models [Eq.~\eqref{eq:1dks} and Eq.~\eqref{eq:1dkpz}] using OTOC as a well-suited diagnostic. The OTOC involves both the spatial spread as well as the temporal growth (or decay) of the initially localized perturbation.  The procedure to compute the classical OTOC is as follows. We initially consider two copies of the height profile: $h_{o}(x,t_{i})$, the \emph{original} copy, and $h_{p}(x,t_i)$, the \emph{perturbed} copy which is generated from the original copy by introducing an infinitesimal local perturbation ($\epsilon$) at initial time $t_{i}$. We then define their difference $\psi(x,t)$ as
\begin{equation}
	\psi( x, t ) 
	:=  \lim_{\epsilon\rightarrow 0}
	\frac{1}{\epsilon} \Big( h_{p}(x,t_i + t ) - h_{o}(x,t_i +t ) \Big), \ 
	t \geqslant 0,
	\label{eq:hdif}
\end{equation}
where $h_o$ and $h_p$ are numerically computed using Eq.~\eqref{eq:1dks} or Eq.~\eqref{eq:1dkpz}, and in the case of KPZ, the two are subject to precisely the same noise $\eta(x,t)$. Then the OTOC, denoted as $D(x, t)$, is defined in terms of $\psi$ in the following manner 
\begin{equation}
    D(x, t) := \langle \left| \psi( x, t ) \right| \rangle ,
    \label{eq:dxt}
\end{equation}
where $ \langle \cdot \rangle $ is average over different initial conditions.  It is to be noted that several works about classical OTOCs define $D(x,t)$ by instead averaging the square of $\psi$. This quantity $D(x,t)$ can be plotted as a ``heat map'' and it encodes the spatial spread and temporal growth or decay of the initial perturbation. The former can be characterized by the \emph{butterfly velocity}, and the latter by the \emph{finite-time Lyapunov exponent} (FTLE) and by the \emph{velocity-dependent Lyapunov exponent} (VDLE).

\begin{center}
\begin{figure}[htbp!]
	\begin{subfigure}{0.5\linewidth}
		\includegraphics[width=1\linewidth]{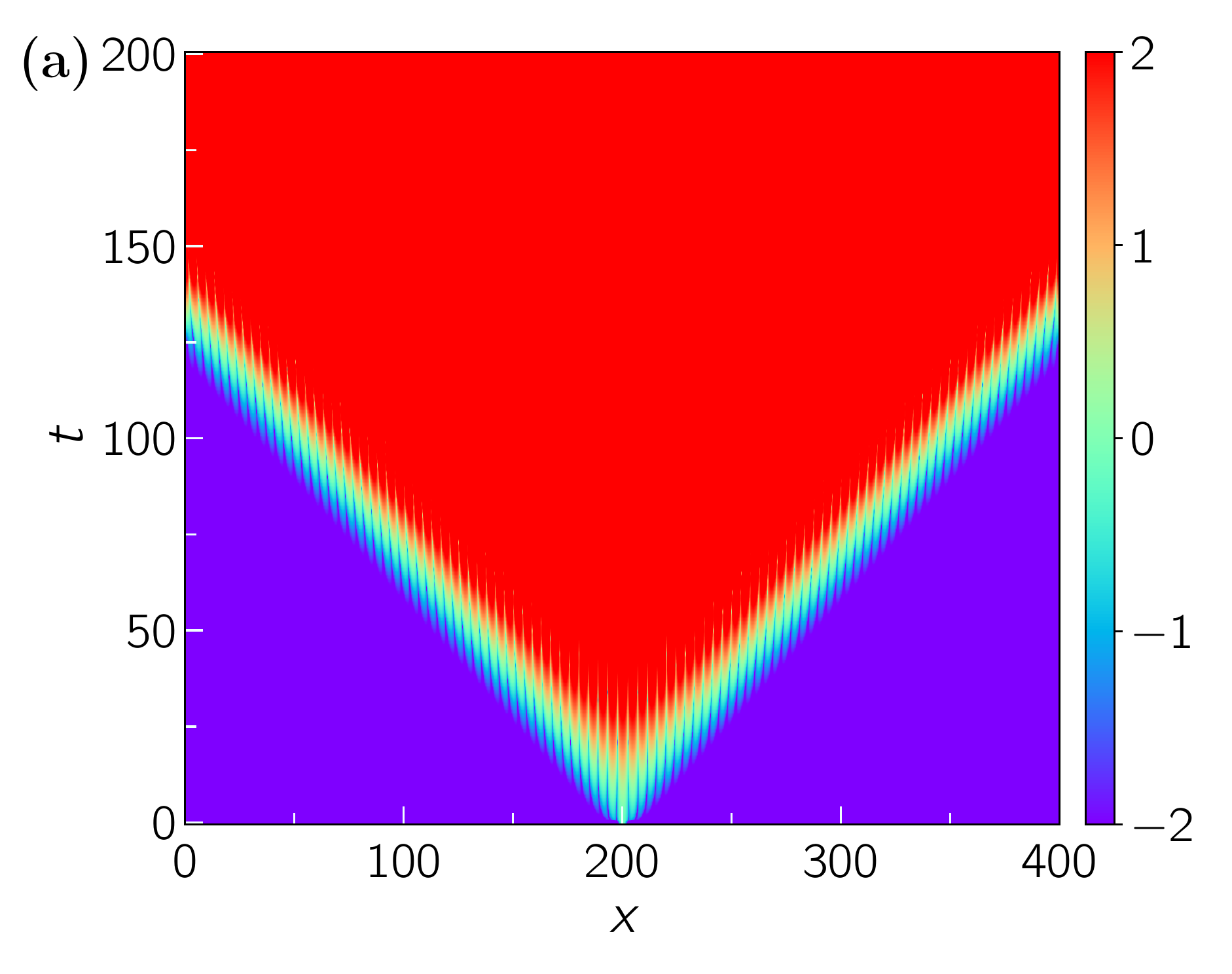}
	\end{subfigure}%
	\begin{subfigure}{0.5\linewidth}
		\includegraphics[width=1\linewidth]{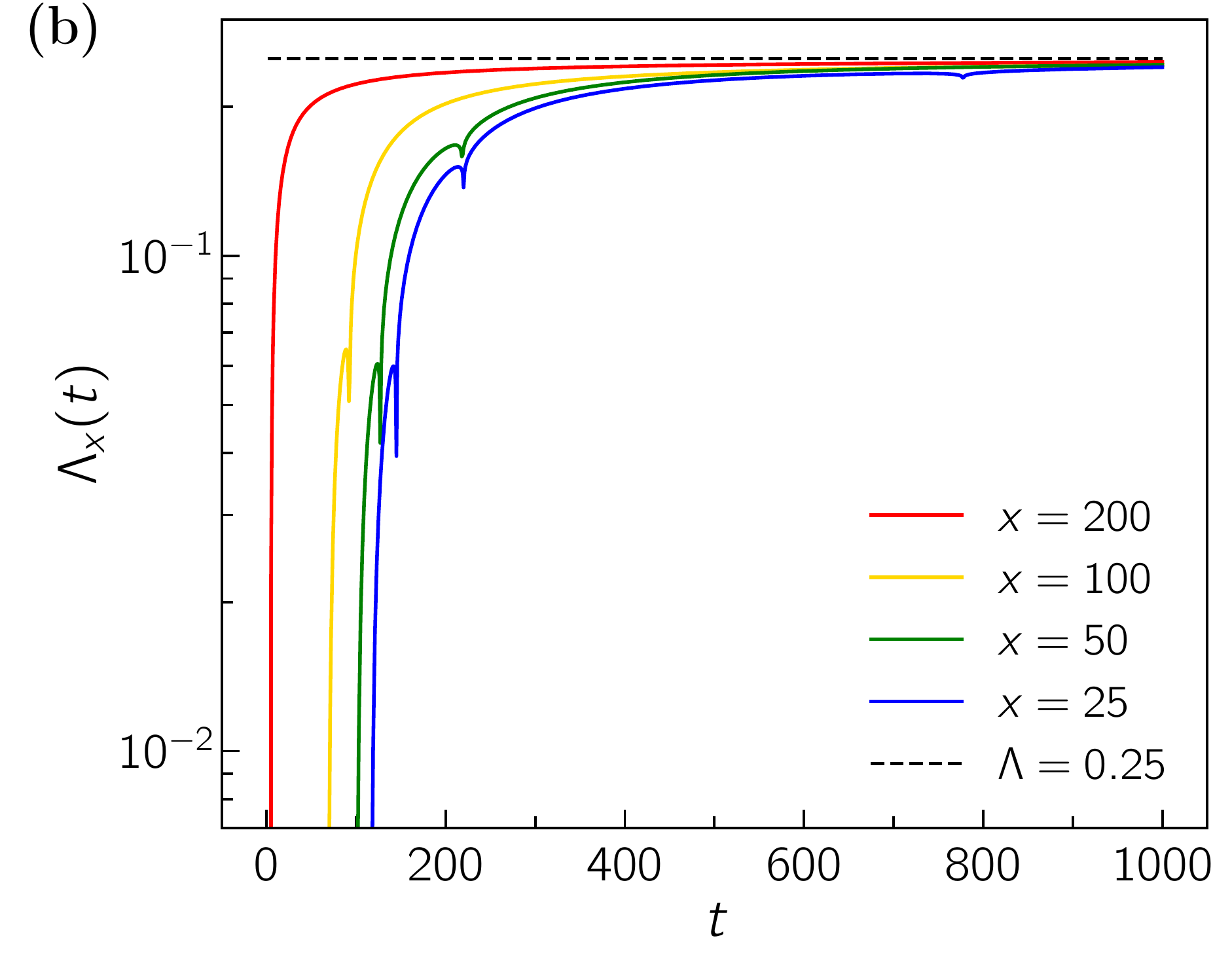}
	\end{subfigure}
	\caption{(Color online) Plots of (a) the heatmap for the OTOC, and (b) FTLE without the nonlinear part in the KS equation in Eq.~\eqref{eq:1dks}. The emergence of a sharp light-cone and non-zero Lyapunov exponent even in the linear model is rooted in the unstable long-wavelength modes. Note that the heat map is for $\log_{10} D$  and location of initial perturbation is at $x=L/2$ (center). Here, we have used $L=400$ and $N=2048$.
    }
    \label{fig:exact-otoc}
\end{figure}	
\end{center}

\begin{center}
\begin{figure}[htbp!]	
    \begin{subfigure}{0.5\linewidth}
		\includegraphics[width=1\linewidth]{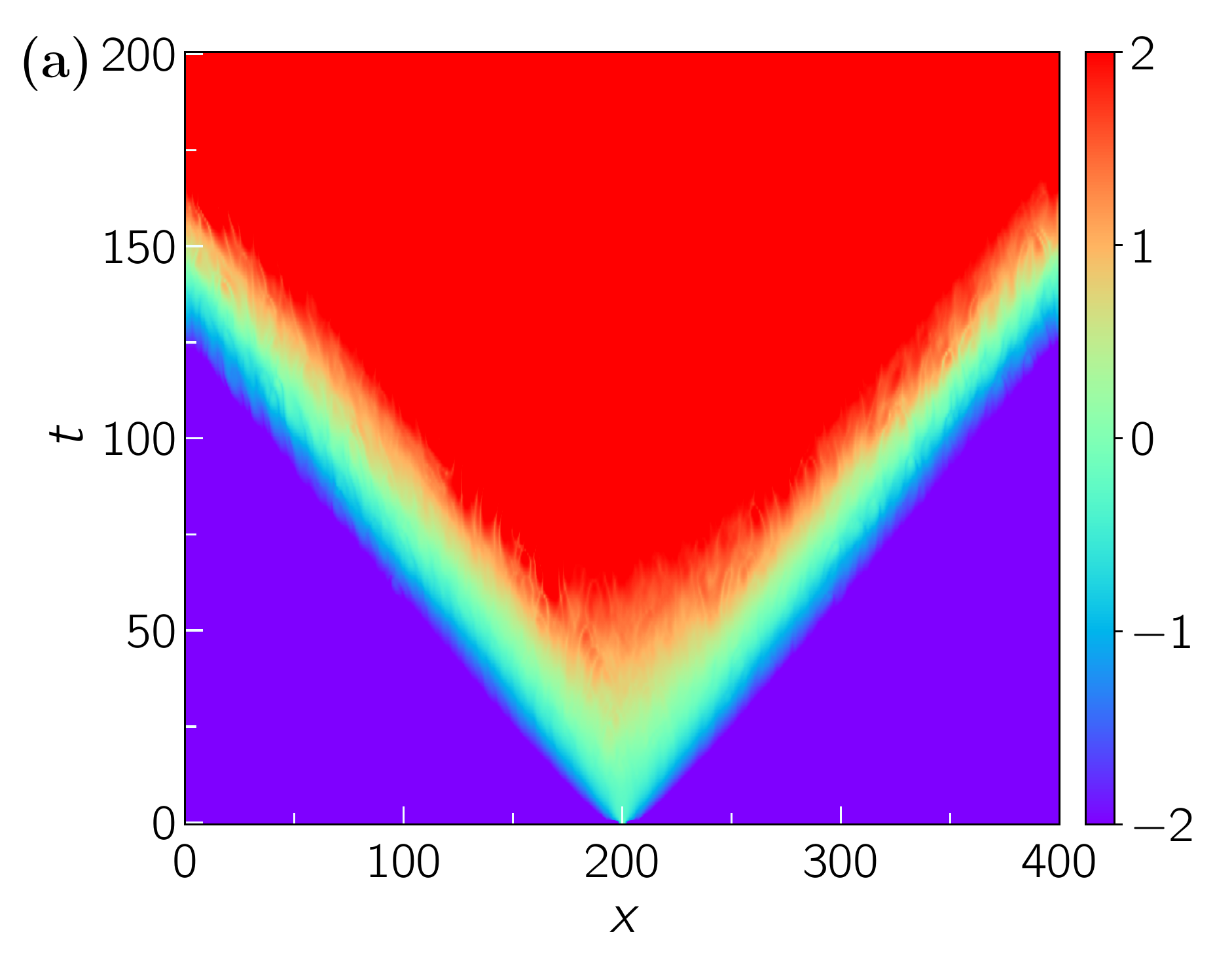}
	\end{subfigure}%
	\begin{subfigure}{0.5\linewidth}
		\includegraphics[width=1\linewidth]{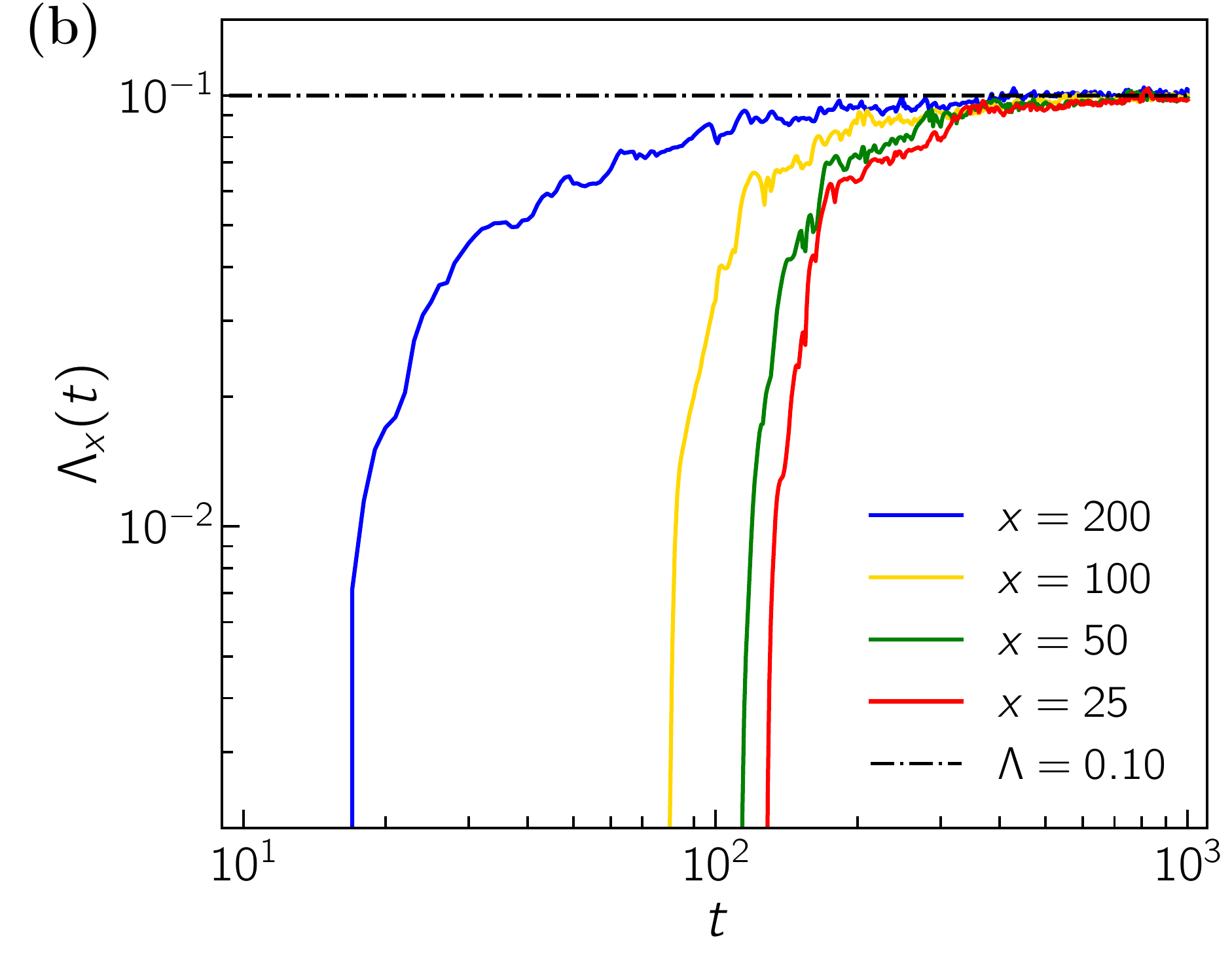}
	\end{subfigure}		
    \caption{(Color online) The plots of (a) the OTOC and (b) the Lyapunov exponent for the fully nonlinear KS equation in Eq.~\eqref{eq:psi-ks} with $L = 400, N=2048$ and $500$ independent simulations. It is worth noting that the value of the butterfly velocity is quite close to the value obtained in the case of linear KS equation. However, the Lyapunov exponents are markedly different and we find that the nonlinear terms substantially reduce the maximum Lyapunov exponent. Similar to Fig~\ref{fig:exact-otoc}, here also the heat map is for $\log_{10} D$  and location of initial perturbation is at $x=L/2$ (center). Perturbation was added at $t_i = 1000$.}
    \label{fig:ks-otoc}
\end{figure}
\end{center}

\begin{center}
    \begin{figure}[htbp!]
	\begin{subfigure}{0.5\linewidth}
		\includegraphics[width=1\linewidth]{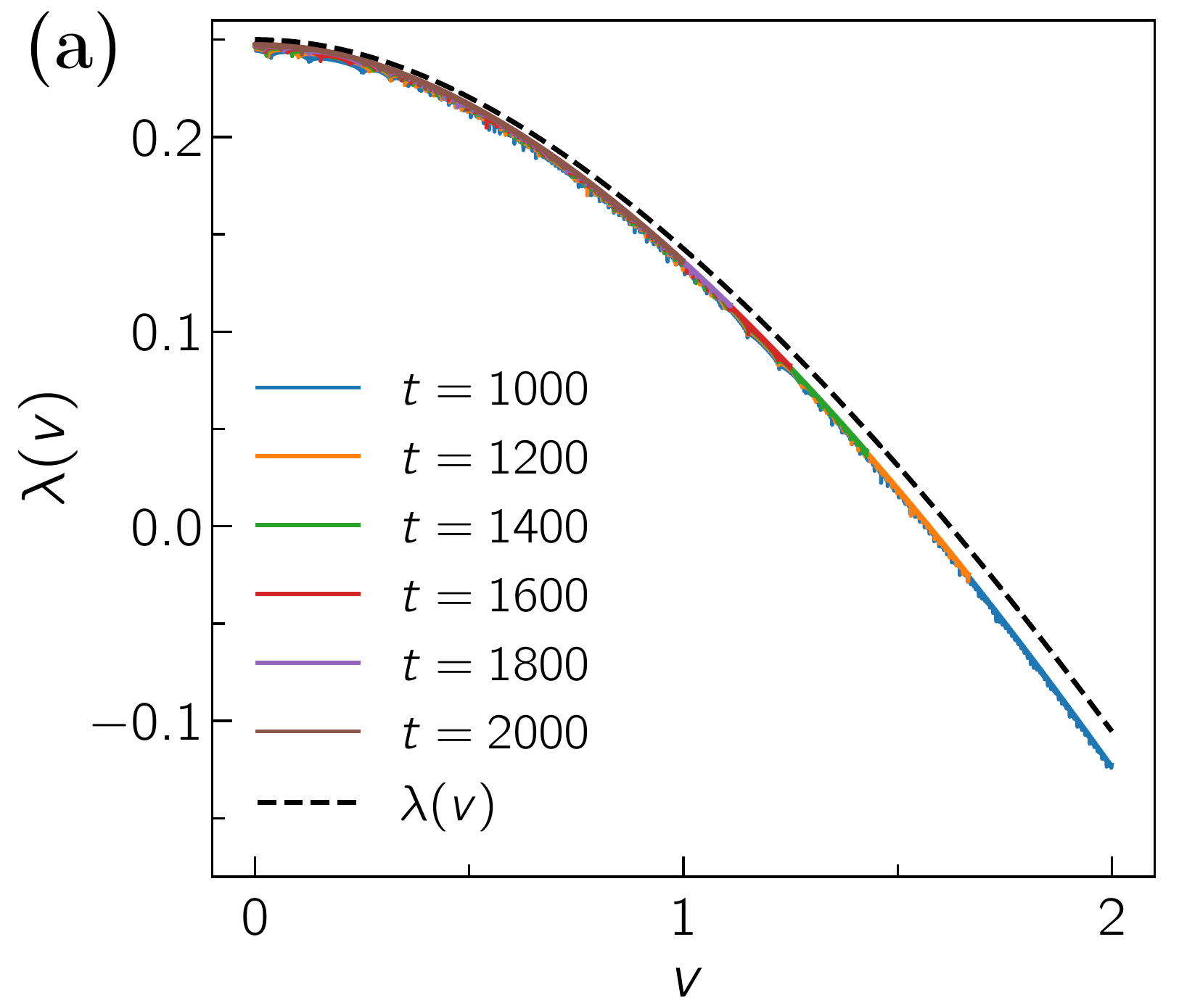}
	\end{subfigure}%
	\begin{subfigure}{0.5\linewidth}
		\includegraphics[width=1.0\linewidth]{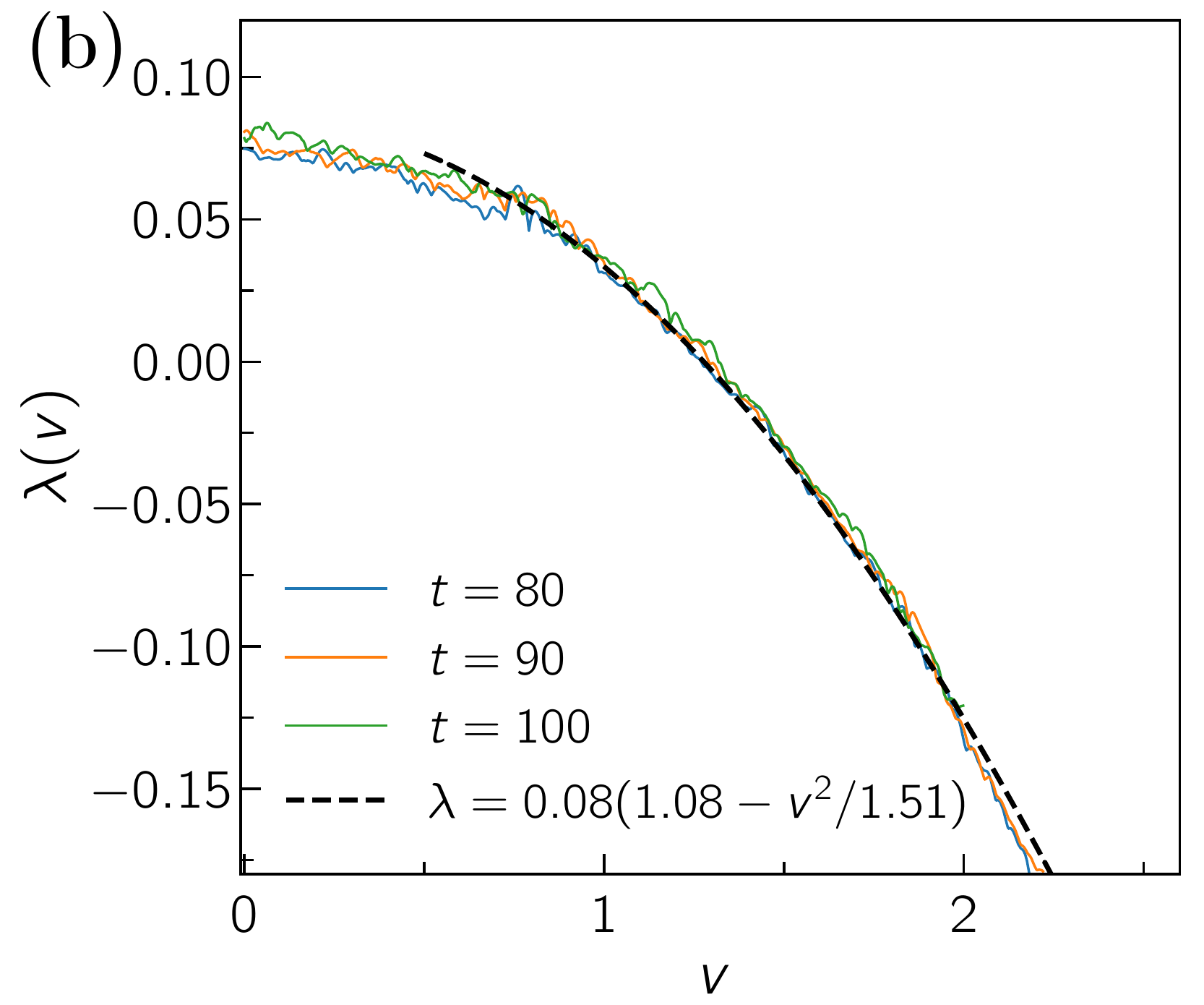}
	\end{subfigure}   
        \caption{(Color online) The plots of the velocity-dependent Lyapunov exponents (VDLE) $\lambda(v)$ for (a) the linear and (b) the fully nonlinear KS equation in Eq.~\eqref{eq:1dks} for $L=4000 \ (N=8192)$ and $L=400 \ (N=2048)$, respectively. Note that in (a), the expression for $\lambda(v)$ is taken from Eq.~\eqref{eq:lamv}. One can notice good agreement between analytical computation and direct numerics in (a). In (b), the black dashed line represents a suitable fit for the VDLE in the fully nonlinear KS equation.}
	\label{fig:ks-vdle}
    \end{figure}
\end{center}

\begin{figure}[htbp!]
    \begin{center}
        \begin{subfigure}{0.5\linewidth}
	\includegraphics[width=1\linewidth]{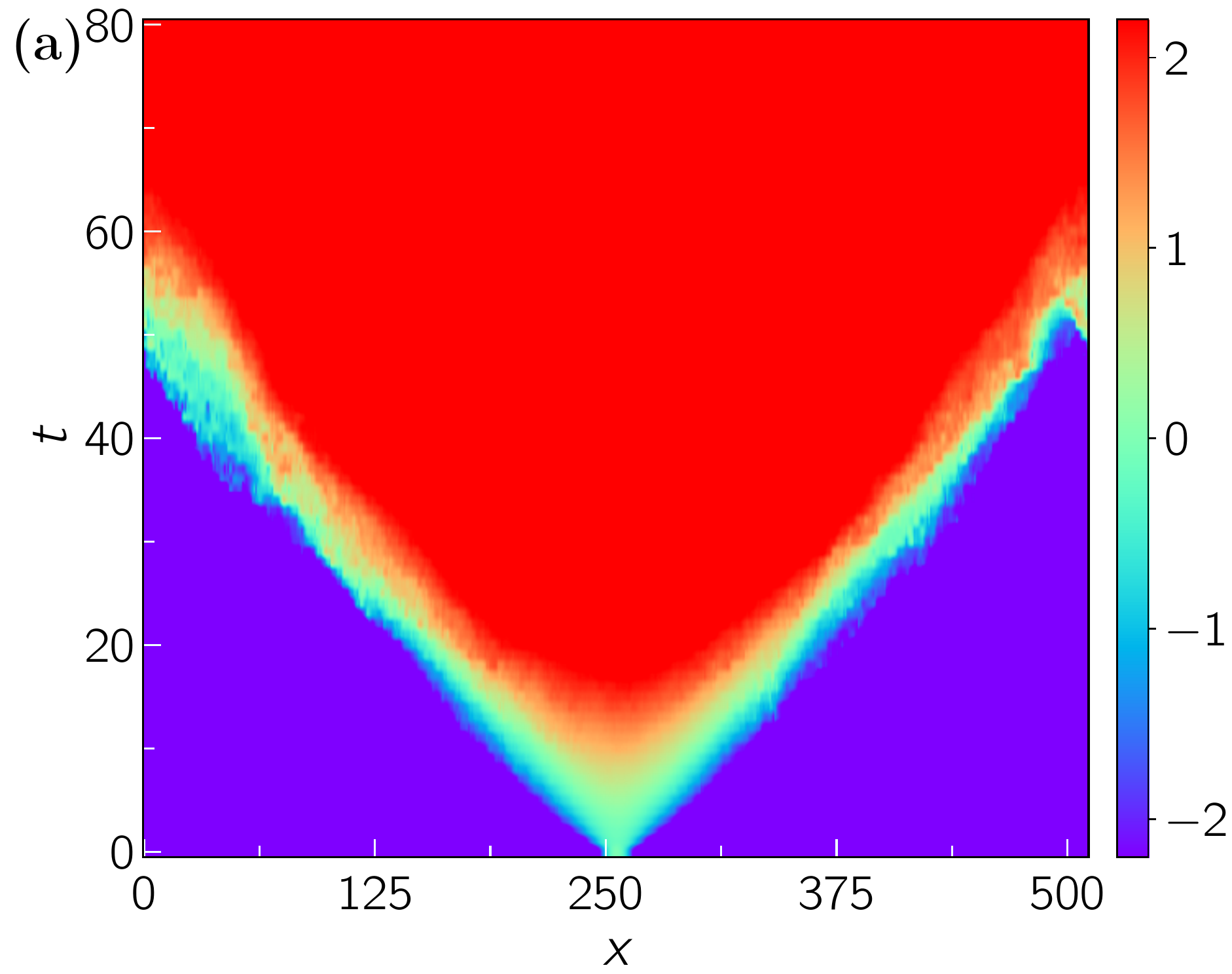}
        \end{subfigure}%
        \begin{subfigure}{0.5\linewidth}
        \includegraphics[width=1\linewidth]{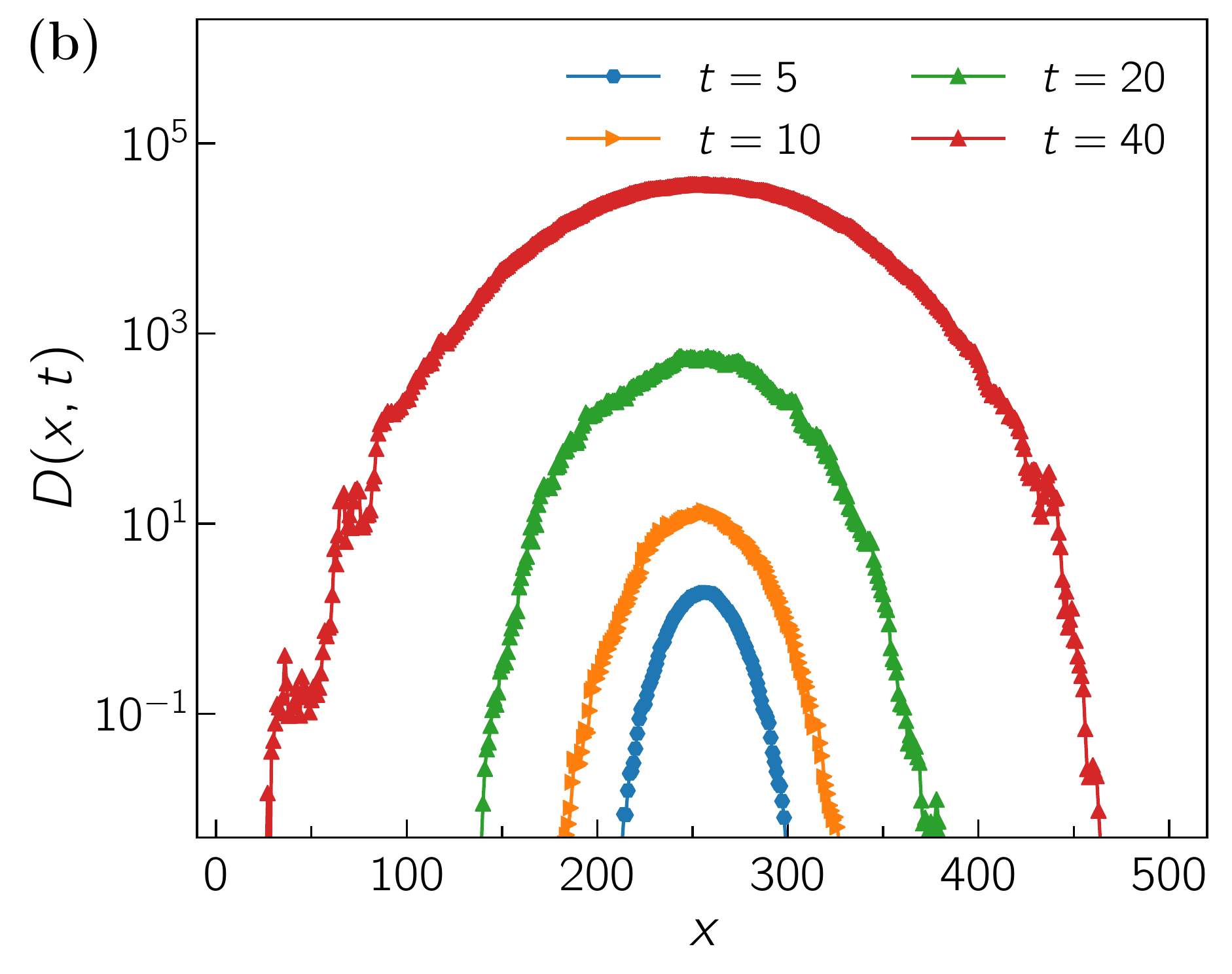}
        \end{subfigure}
    \end{center}
    \caption{(Color online) (Left) The OTOC for the KPZ equation [see Eq.~\eqref{eq:1dkpz} and Eq.~\eqref{eq:hdif}] for $g=8$ using the Lam-Shin finite-difference method (see Supplemental Material \cite{2023-supp}) with $L=512, \epsilon = 10^{-5}, N=512$ with $w=4$. Total number of independent simulations is $14900$. Note that the heat map is for $\log_{10} D$, and location of initial perturbation is at $x=L/2$ (center). Perturbation was added at $t_i = 500$. (Right)  Behaviour of the OTOC $D(x,t)$ in Eq.~\eqref{eq:dxt} as a function of $x$ for various time snapshots $t=5, 10, 20, 40$. Both the left and the right moving fronts show a ballistic propagation. The slowdown from exponential temporal growth during the time window $t=20$ to $t=40$ can be attributed to the finiteness of $\epsilon$.}
    \label{fig:kpz-otoc}
\end{figure}

\textit{Summary of findings. --} The key findings of our investigation are as follows.
\begin{enumerate}
    \item  For the KS equation, we observe a sharp light-cone in the OTOC even in the linearized case (Fig.~\ref{fig:exact-otoc}), i.e. neglecting the term $ - \brac{ \partial_x h }^2 / 2 $ in Eq.~\eqref{eq:1dks}, demonstrating that chaos due to nonlinearity is not needed to produce such a ballistically spreading OTOC. We compute the exact expression of OTOC in this case and unravel the interplay between unstable 
    modes and dissipation. Using the method of steepest descent, we extract the values of butterfly velocity and Lyapunov exponents. We then use extensive numerical simulations to compute the OTOC of the fully nonlinear (Fig.~\ref{fig:ks-otoc}) KS equation Eq.~\eqref{eq:1dks}. The velocity-dependent Lyapunov exponents have been studied in both the linear and the fully nonlinear KS equation (Fig.~\ref{fig:ks-vdle}).


    \item We investigate the OTOC in the 1D KPZ equation (Fig.~\ref{fig:kpz-otoc}) using the discretization scheme provided in Ref.~\onlinecite{1998-lam-shin}. Following this numerical discretization scheme, we observe a conventional light-cone in the OTOC for the KPZ equation. It is important to note that, even though certain statistical properties (such as scaling, spatiotemporal correlations and height distributions) related to the 1D KPZ equation are correctly reproduced by the method in Ref.~\onlinecite{1998-lam-shin}, the chaotic behaviour that we have characterized using the OTOC is expected to be true only for the discretized KPZ equation \cite{2023-barraquand-doussal-priv-comm}. 
    
\end{enumerate}

\textit{Results for the KS equation.--} To study the OTOC in the KS equation, we focus on $\psi(x, t)$ given in Eq.~\eqref{eq:hdif}. The KS equation in Eq.~\eqref{eq:1dks} in the limit of infinitesimally small perturbation ($\epsilon \rightarrow 0$) reads
\begin{equation}
	\partial_{t} \psi 
	= 
	- \partial^{2}_{x} \psi - \partial^{4}_{x} \psi  
	- \partial_{x} h_{o} \, \partial_{x} \psi  \ .
    \label{eq:psi-ks}
\end{equation}
Eq.~\eqref{eq:psi-ks} is a linear equation in $\psi$ with coefficients dictated by the evolution of the height field $h_o(x,t)$ evolving according to Eq.~\eqref{eq:1dks}. We choose the following initial condition for $\psi$ where the two copies of the height profiles differ only near $x=L/2$ (centre) at $t=0$ in the following manner: 
\begin{equation}
	\psi( x ,0) 
	= 	
	\exp \! \left[  - \frac{ (x-L/2)^2 }{ w^2 } \right],
	\label{eq:phi-ic}
\end{equation}
where $0<w<<L$ determines the width of the Gaussian perturbation. We set this width to be $w=1$ for studying OTOC in the KS equation.

The lateral extent of the light-cone given by Eq.~\eqref{eq:dxt} gives the spatial spread of the initial perturbation thereby yielding the \emph{butterfly velocity} $v_b$ which characterizes  the speed with which the boundary of the light-cone, defined by $D(x,t)=1$, moves. On the other hand to understand the temporal growth or decay at some fixed spatial point $x$, we define the \emph{finite-time Lyapunov exponent} (FTLE) as 
\begin{equation}
 \Lambda_{x}( t ) :=  \frac{\ln D (x, t )} {t}. 
 \label{eq:ftle}
\end{equation}

The velocity-dependent Lyapunov exponent (VDLE) can be defined as 
\begin{equation}
    \lim_{t \rightarrow \infty} \frac{\ln D(x=vt,t)}{ t} = \lim_{t \rightarrow \infty} \Lambda_{x=vt}(t) = \lambda (v).
    \label{eq:vdle}
\end{equation}
For these systems, the maximal Lyapunov exponent is $\lambda(v=0)$. 


Before presenting extensive numerical results of Eq.~\eqref{eq:psi-ks} which will give the OTOC for the fully nonlinear KS equation, we will first present some results for the linearized KS equation. This is equivalent to studying Eq.~\eqref{eq:psi-ks} without the last term.  A saddle-point analysis of the late time behavior can be done for this linearized model.  

In the Fourier space, ignoring the last term in Eq.~\eqref{eq:psi-ks}, the wavenumber-$k$ mode $\widetilde{\psi}_{k}( t )$ obeys 
\begin{equation}
        \partial_{t} \widetilde{\psi}_{k_n}  = r_{k_n} \widetilde{\psi}_{k_n}
        \label{eq:psi-modes}
\end{equation}
where
\begin{equation}
    r_{k_n} = k_n^{2} - k_n^{4}, \text{ and } \quad \widetilde{\psi}_{k_n} = \frac{1}{N} \sum_{m=0}^{N} \psi \Big( \tfrac{m L}{N},t \Big ) e^{ -\frac{ i k_n m L}{N} } ,
\end{equation}
with $k_n =  2 \pi n / L,  n \in \mathbb{Z} $. The solution of Eq.~\eqref{eq:psi-modes} is easily found to be 
\begin{equation}
    \widetilde{\psi}_{k}(t) = \widetilde{\psi}_{k}(0) \exp\!\left( r_{k} t \right).
    \label{eq:psikt}
\end{equation}
Note that the Fourier modes grow if $r_{k}>0$. Thus the modes satisfying $0<k_n^2<1$ grow with time. For a discretized system with $N$ equispaced gridpoints, in the real space, the difference $\psi(x,t)$ is
\begin{equation}
	\psi( x, t ) 
	\approx 
	\widetilde{\psi}_{0}(0) + 2 \sum_{ n=1 }^{ \frac{N}{2} } e^{ r_{k_{n}} t }  \text{Re} \! \left[  \widetilde{\psi}_{k_{n}}(0) e^{i k_{n} x} \right] , 
	\label{eq:sum}
\end{equation}
where we recall that $k_{n} = {2 \pi n}/{L}$ are the modes in the discretized system.

\begin{center}
\begin{figure}[htbp!]	
    \includegraphics[width=1\linewidth]{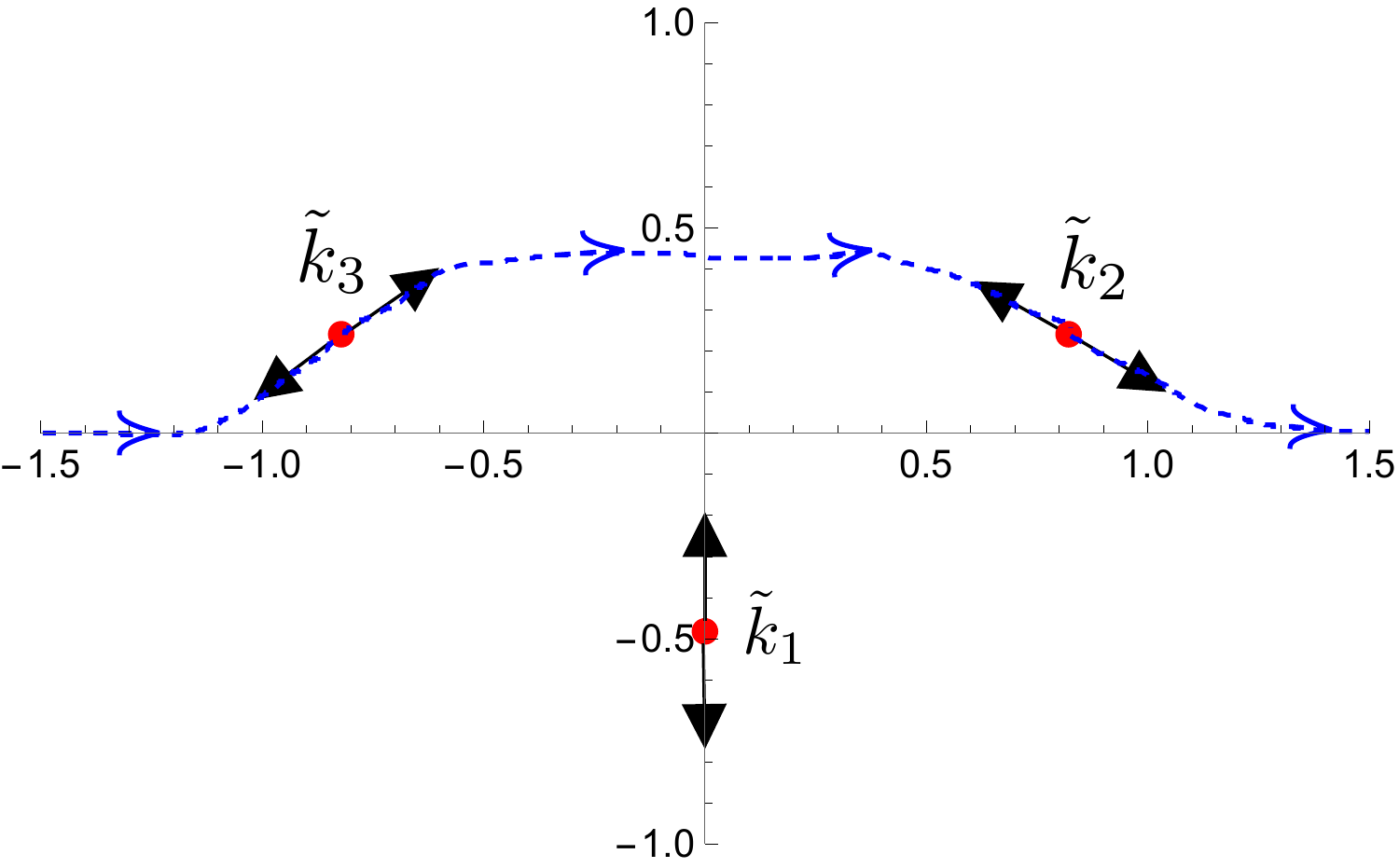}  
    \caption{(Color online) A schematic diagram of the deformed contour needed to perform the integration. The red dots show the three roots for an arbitrary chosen sample value of $v=\sqrt{2}$ for the saddle point analysis. The method in Ref.~\onlinecite{2003-ablowitz-fokas} is adapted here. }
    \label{fig:schematic-saddle-roots}
\end{figure}
\end{center}

In the Fourier space, the initial condition for $\psi(x, t )$ given in Eq.~\eqref{eq:phi-ic} becomes
\begin{equation}
	\widetilde{\psi}_{k_{n}}(0) 
	= \frac{w\sqrt{\pi}}{L}        e^{ - \frac{ik_nL}{2} - \frac{k_n^2w^2}{4}}.
\end{equation}
Then, the solution at time $t>0$ is given by [using Eq.~\eqref{eq:sum}]
\begin{equation}
	\psi(x, t ) 
	\approx 
        \frac{w\sqrt{\pi}}{L} \sum_{n=-N/2}^{N/2} e^{ r_{k_n}t + ik_n \overline{x} - \frac{k_n^2 w^2}{4}} ,
\end{equation}
where $\bar{x} = x - L/2$. Thus the corresponding OTOC is given by 
\begin{equation}
    \begin{aligned}
	D(x, t ) 
        \approx
        &\frac{w\sqrt{\pi}}{L}  \Bigg| 1 + 2 \sum_{n=1}^{N/2} \cos(k \overline{x}) \ e^{r_{k_n}t - \frac{k_n^2 w^2}{4}} \Bigg| .
    \end{aligned}
	\label{eq:exact-dxt}
\end{equation}
Note that Eq.~\eqref{eq:exact-dxt} is exact and plotted in Fig.~\ref{fig:exact-otoc}. We note from Fig.~\ref{fig:exact-otoc} that the butterfly velocity turns out to be $v_b \approx 1.6$ and the maximum Lyapunov exponent is $\Lambda \approx 0.25$. Interestingly these values can be
extracted by analytical analysis of Eq.~\eqref{eq:exact-dxt} via the method of steepest descent which we will present below.

We convert the sum in Eq.~\eqref{eq:exact-dxt} into an integration which yields
\begin{equation}
    D(x,t) = \frac{w}{2\sqrt{\pi}}\Big| \int_{-\infty}^{\infty} \mathrm{d}k \ e^{ tg(k)}e^{ -k^2w^2/4}\Big|
    \label{eq:dxt-integ}
\end{equation}
We consider the function in the exponent of the integrand of Eq.~\eqref{eq:dxt-integ},
\begin{equation}
g(k) = i k \frac{\overline{x}}{t} +  (k^2 - k^4). 
\end{equation}
The first and second derivatives of $g(k)$ are respectively given by 
\begin{equation}
    g'(k) = i\frac{\overline{x}}{t} + (2k- 4k^3 ) , \ \ g''(k) = 2 - 12 k^2 .
\end{equation}
Setting $g'(k) = 0$ and solving for $k$, we find three saddle points:
\begin{equation}
   \tilde{k}_m (v)= \frac{i}{\sqrt{2}} \brac{ - \omega^{m-1} z(v) + \frac{1}{3\ \omega^{m-1} z(v)} }, \ m=1, 2, 3,
\end{equation}
where
\begin{equation}
    \omega = \frac{-1 + i\sqrt{3}}{2}, \ \text{ and } \
    z(v) = 
    \sqrt[3]{ \frac{v}{2\sqrt{2}} + \sqrt{ \frac{v^2}{8} + \frac{1}{27}}}
    \label{eq:zv}
\end{equation}
with $v=\bar{x}/t$. Note that the real parts of $\tilde{k}_2$ and $\tilde{k}_3$ have opposite signs but same absolute values, whereas the imaginary parts of these solutions are same. 
These three roots are shown in Fig.~\ref{fig:schematic-saddle-roots} for an arbitrary chosen sample value of $v=\sqrt{2}$.  We deform our integral to pass through the two saddle points at $k=\tilde{k}_2, \tilde{k}_3$.  We need to evaluate $g''(k)$ at these points:
\begin{equation}
    g''(k) = 
        - \brac{ 2 + 3z^2 + \frac{1}{3z^2} \pm i \frac{ 9 z^4 -1}{\sqrt{3} z^2} },  ~k=\tilde{k}_2, \tilde{k}_3,
\end{equation}
where $z$ is given in Eq.~\eqref{eq:zv} and we omit the argument $v$ for the sake of brevity. We adopt the procedure in Ref.~\onlinecite{2003-ablowitz-fokas} for the method of steepest descent to evaluate $D(x,t)$ in Eq.~\eqref{eq:dxt-integ}. Recall that we have two stationary points  ($\tilde{k}_2, \tilde{k}_3$) along the contour as observed above and the value of $D(x,t)$ in the limit of long time is sum of the contributions from these points.

To reduce our problem to a form adaptable to the procedure in Ref.~\onlinecite{2003-ablowitz-fokas}, note that 
\begin{eqnarray}
    g(\tilde{k}_2) &=&  \Big[ \frac{v}{\sqrt{2}} \brac{\omega \, z - \frac{1}{3\, \omega \,z } } - \frac{1}{2} \brac{\omega z - \frac{1}{3 \,\omega\, z } }^2 \nonumber \\
    && - \frac{1}{4} \brac{\omega\, z - \frac{1}{3\, \omega \,z } }^4 \Big] .
    \label{eq:gk2}
\end{eqnarray}
Also it is easy to see that the directions of steepest descent for $\tilde{k}_2$ are given by  
\begin{equation}
\label{eq:theta2}
    \theta_2 = - \frac{1}{2}\text{Arg}[g''(\tilde{k}_2)] + \frac{3\pi}{2}, 
\end{equation}
where our notation is such that an angle $\theta=0$ corresponds to the positive real axis. We deform our contour (Fig.~\ref{fig:schematic-saddle-roots}) at $\tilde{k}_2$ along the direction dictated by $\theta_2$ given in Eq.~\eqref{eq:theta2}.
From Eq.~\eqref{eq:theta2}, it turns out that $\theta_2$ lies in the fourth quardrant, $11\pi/6<\theta_2<2\pi$.
Similarly, the contour at $\tilde{k}_3$ is deformed as per $\theta_3$ given by
\begin{equation}
\label{eq:theta3}
    \theta_3 = - \frac{1}{2}\text{Arg}[g''(\tilde{k}_3)] + \frac{\pi}{2}, 
\end{equation}
and using Eq.~\eqref{eq:theta3}, it turns out that $\theta_3$ lies in the first quadrant, $0<\theta_2<\pi/6$. We need to integrate along these directions and add the contributions. Therefore, adapting Ref.~\onlinecite{2003-ablowitz-fokas} we find that the contribution from the saddle point $\tilde{k}_2$ is given by
\begin{equation}
    D_2 (x,t) = 
    \frac{\sqrt{2 \pi } }{ \sqrt{ t |g''(\tilde{k}_2)|} } 
    e^{ i \theta_2 + t g(\tilde{k}_2)- (\tilde{k}_2^2w^2/4)} ~,
\end{equation}
while the contribution from $\tilde{k}_3$ is the 
complex conjugate of this. 
Combining the contributions, we get the following result:
\begin{equation}
    D(x,t) \simeq
    \frac{2 \sqrt{2 \pi } }{ \sqrt{ t |g''(\tilde{k}_2)|} } 
    \Big| \text{Re} \left[ e^{ i \theta_2 + t g(\tilde{k}_2)-(\tilde{k}_2^2w^2/4)}\right] \Big|.
    \label{eq:otoc-analytical}
\end{equation}
The velocity-dependent Lyapunov exponent is then given by the exponential growth (or decay) of this with time:
\begin{equation}
    \lambda(v) = \text{Re} \left[ g(\tilde{k}_2) \right] ~.
    \label{eq:lambda_v}
\end{equation}
Using Eq.~\eqref{eq:gk2}, $\lambda(v)$ in Eq.~\eqref{eq:lambda_v} takes the form,
\begin{eqnarray}
\lambda(v) &=& \frac{z^4}{8}+\frac{z^2}{12}-\frac{v \,z}{2 \sqrt{2}}+\frac{v}{6 \sqrt{2} \,z}\nonumber \\&+& \frac{1}{108 \,z^2}+\frac{1}{648 \,z^4}+\frac{1}{6} ~,
\label{eq:lamv}
\end{eqnarray}
where we recall that $z(v)$ is given by Eq.~\eqref{eq:zv}. Although Eq.~\eqref{eq:lamv} is rather cumbersome, it turns out that $\lambda(v=0) = 0.25$ which is the maximal Lyapunov exponent. The butterfly velocity can be extracted by solving for $\lambda(v_b) =0$. It turns out that the butterfly velocity $v_b$ extracted in this way is $v_b \approx 1.62$. Therefore Lyapunov exponent and butterfly velocity extracted are in good agreement with the numerically obtained values. 

Since we see from these analytical calculations that $D(x,t)$ goes as $ \sim e^{t \lambda(v)}$, we thus obtain a ``light-cone'' behavior for this OTOC that is qualitatively the same as is seen in many-body chaos.  But this is appearing in a linear equation due to its linear instability.

Having discussed the saddle-point analysis and the linear KS equation, we now present results for the fully nonlinear KS equation using extensive numerics. 
As shown in Fig.~\ref{fig:ks-otoc}, we observe a distinct light-cone in the heatmap of the OTOC and also temporal growth in chaos unravelled by the finite-time Lyapunov exponent. Interestingly, we find the value of the butterfly velocity ($v_b \approx 1.50$) is changed very little 
when nonlinearity is included. On the other hand, remarkably, the maximum Lyapunov exponent $\lambda(v=0)$ shows a large decrease when nonlinearity is included. In the linear KS equation, $\lambda(v=0)$ is set by the most unstable linear modes. 
However, in the nonlinear case, the interaction term strongly couples all the linear modes; apparently this causes the maximum Lyapunov exponent to be more like an average over many of the linear modes, and thus much smaller than that of the most unstable linear mode.  This behavior is in contrast to adding a nonlinearity to a linearly {\it stable} system, where the nonlinearity causes chaos and thus an {\it increase} of the maximum Lyapunov exponent.  
We also show the velocity-dependent Lyapunov exponent both for the linear and fully nonlinear KS equation in Fig.~\ref{fig:ks-vdle}.

It is important to recall that there have been studies \cite{2020-roy-pandit} showing deep connection between the KS and KPZ equations. In particular, Tracy-Widom and Baik-Rains distributions which were observed for the 1D KPZ equation earlier \cite{2000-prahofer-spohn} were also shown to occur in the KS equation. One naturally wonders whether there is such a connection in the OTOC as well. 
We next discuss the OTOC and related quantities in the KPZ equation under a lattice discretization and numerical scheme given in Ref.~\onlinecite{1998-lam-shin}.

For the KPZ case, the equation obeyed by the difference field $\psi$ given in Eq.~\eqref{eq:hdif} (in the limit of infinitesimally small
perturbation) is
\begin{equation}
    \partial_{t} \psi  =  \partial^{2}_{x} \psi + 2g \partial_x \psi \partial_x h_o,  
    \label{eq:psi-kpz}
\end{equation}
where recall that $h_o$ is the original height field satisfying KPZ equation given in Eq.~\eqref{eq:1dkpz}. Note that albeit Eq.~\eqref{eq:psi-kpz} is linear in $\psi$, the presence of the stochastic field $h_o(x,t)$ is what gives rise to sharp lightcones and related features of course, assuming the numerical discretization. In order to study the OTOC for the KPZ equation, we employ the Lam-Shin finite difference method \cite{1998-lam-shin} and we resort to the method of two copies in Eq.~\eqref{eq:hdif}. We describe the Lam-Shin finite-difference method in Supplemental Material~\cite{2023-supp}. In Fig.~\ref{fig:kpz-otoc}, using extensive numerics we present results for the light-cone which is characterized by a butterfly velocity $v_b \approx 4.8 $ and FTLE $\Lambda \approx 0.32 $ for nonlinearity strength $g=8$. 
As mentioned earlier, our results are valid only in the discretized KPZ equation and will not hold in the strictly continuum KPZ equation \cite{2023-barraquand-doussal-priv-comm}. Therefore, despite the established deep connections in the long time and large system size limit between the continuum KS and continuum KPZ equation, it is important to keep in mind that certain quantities such as OTOC are expected to be strikingly different.

\textit{Conclusions and Outlook. --} We have studied the spatiotemporal spread of an initial localized perturbation using the OTOC in the 1D Kuramoto-Sivashinsky (KS) equation. 
This is a deterministic nonlinear differential equation with unstable long-wavelength modes stabilized by nonlinear terms. Via extensive numerical simulations we have characterized spatial spread and temporal growth in the KS equation in the continuum limit. We provide an analytical insight for the linearized KS equation which has a unique property of hosting a well-defined light-cone structure even in the linear regime. The role of the unstable long-wavelength modes in the linearized KS equation has been understood by a saddle-point analysis. We also provide results for the KPZ equation under a numerical discretization scheme described in Ref.~\onlinecite{1998-lam-shin}. However, in the truly continuum limit, the KPZ equation is not expected to show spatiotemporal chaos \cite{2023-barraquand-doussal-priv-comm}. The positive largest Lyapunov exponent in the KS equation has the linearly unstable long-wavelength modes as its source, while for the (discretized) KPZ equation there are no linearly unstable modes and the chaos appears to be ``sourced'' at the scale of the numerical discretization.

Our work demonstrates that the KS equation is an excellent platform for studying chaos in spatially continuum systems. It will be interesting to explore spatiotemporal chaos in multi-component systems \cite{1992-ertas-kardar, 1993-ertas-kardar, 1999-kliakhandler, 2000-tasev--parlitz, 2001-das--ramaswamy, 2013-ferrai--spohn, 2015-spohn-stoltz, 2022-hayashi-arxiv} where one can study chaos in different species. Given that many physical systems fall into the 1D KPZ universality class \cite{1995-healy-zhang, 2015-healy-takeuchi, 2016-spohn-arxiv, 2018-takeuchi, 2020-spohn}, our findings should hold for such systems of both experimental and theoretical interest.  \\

\textit{Acknowledgements. --}
We would like to acknowledge Guillaume Barraquand and Pierre Le Doussal for very useful discussions. M.K. would like to acknowledge support from the project 6004-1 of the Indo-French Centre for the Pro- motion of Advanced Research (IFCPAR), Ramanujan Fellowship (SB/S2/RJN-114/2016), SERB Early Career Research Award (ECR/2018/002085) and SERB Matrics Grant (MTR/2019/001101) from the Science and Engineering Research Board (SERB), Department of Science and Technology (DST), Government of India. M.K. acknowledges support of the Department of Atomic Energy, Government of India, under Project No. 19P1112RD.  D.A.H. was supported in part by (USA) NSF QLCI grant OMA-2120757.

\medskip

\bibliographystyle{apsrev4-1}
\bibliography{ks-otoc-refs.bib}

\end{document}


\newcommand{\titlename}{\underline{\textsc{Supplemental material}}\\ \bigskip Out-of-time-ordered correlator in the one-dimensional Kuramoto-Sivashinsky and Kardar-Parisi-Zhang equations}

\title[]{\titlename}

\author{Dipankar Roy}
\email[ ]{dipankar.roy@icts.res.in}
\affiliation{International Centre for Theoretical Sciences, Tata Institute of Fundamental Research,
	Bangalore 560089, India}

\author{David A. Huse}
\email[ ]{huse@princeton.edu}
\affiliation{Physics Department, Princeton University, Princeton, NJ, 08544, USA}

\author{Manas Kulkarni}
\email[ ]{manas.kulkarni@icts.res.in}
\affiliation{International Centre for Theoretical Sciences, Tata Institute of Fundamental Research,
	Bangalore 560089, India}
\date{\today}


\maketitle

\tableofcontents 



\section{Numerical methods for the KPZ equation \label{sec:app-nmkpz}}
Here we discuss the numerical method we employ for solving the KPZ equation. The Lam-Shin method \cite{1998-lam-shin} is a finite-difference technique where central difference is used for the second-derivative term and the nonlinear term is handled with a modified difference term adapted for the 1D KPZ equation. The height profile $h_n$ at the $n$-th grid point (assuming periodic boundary conditions) satisfies
\begin{equation}
\frac{ \textrm{d} h_{n} }{ \textrm{d} t} = C_{n} + g N_n + \xi_n ,
\label{eq:ls-kpz}
\end{equation}
where 
\begin{equation}
\begin{aligned}
C_n  & = h_{n+1} + h_{n-1} - 2 h_{n}, \\
N_n  & = \frac{1}{3} \big[ \brac{ h_{n+1} - h_{n} }^{2} + \brac{ h_{n+1} - h_{n} }\brac{ h_{n} - h_{n-1} } \\
& \quad \  + \brac{ h_{n} - h_{n-1} }^{2} \big] .
\end{aligned}
\end{equation}
Note that here we set $\Delta x = L/ N $ to $1$ \cite{1998-lam-shin}. Thus the height $h_{n}$ in the Lam-Shin numerical scheme is directly coupled only to the nearest neighbours. With this discretization shown in Eq.~\eqref{eq:ls-kpz}, we use Euler-Maruyama method for time-marching \cite{1998-lam-shin, 1992-kloeden-platen}.

\bibliographystyle{apsrev4-1}
\bibliography{ks-otoc-refs.bib}